\pdfoutput=1
\documentclass[twocolumn,epjc3]{svjour3}
\smartqed
\usepackage[numbers,sort&compress]{natbib}
\usepackage{graphicx}
\usepackage{amsmath,amssymb,amsfonts,bm}
\usepackage{booktabs,multirow,xcolor,url,placeins,needspace,stfloats,float}
\usepackage[hidelinks]{hyperref}
\journalname{Eur. Phys. J. C}
\begin{document}
\makeatletter\@fleqnfalse\makeatother
\title{Initial-state control of vorticity--shear competition in longitudinal $\Lambda$ polarization in Pb--Pb collisions at $\sqrt{s_{\mathrm{NN}}}=5.36\,\mathrm{TeV}$}
\titlerunning{Initial-state control of longitudinal $\Lambda$ polarization}
\author{Simin Wu\thanksref{equal} \and Leyao Lin\thanksref{equal} \and Yilong Xie\thanksref{corresponding}}
\institute{School of Mathematics and Physics, China University of Geosciences (Wuhan), Lumo Road 388, 430074 Wuhan, China\\
\email{xieyl@cug.edu.cn}}
\thankstext{equal}{Simin Wu and Leyao Lin contributed equally to this work.}
\thankstext{corresponding}{Corresponding author.}
\date{}
\maketitle
\begin{abstract}
We investigate how the transverse initial energy-density distribution controls the second sine harmonic of longitudinal $\Lambda$ polarization in Pb--Pb collisions at $\sqrt{s_{\mathrm{NN}}}=5.36\,\mathrm{TeV}$. We use (3+1)-dimensional Particle-In-Cell Relativistic ideal hydrodynamics and the Becattini--Buzzegoli--Palermo isothermal local-equilibrium prescription. Across the twelve selected initial states, the kinematic-vorticity contribution $P_{z,s2}^{\omega}$ is negative, whereas the kinematic-shear contribution $P_{z,s2}^{\Xi}$ is positive. Changing the initial energy-density normalization $e_{\mathrm{peak}}$, transverse smoothing width $\sigma_\perp$, or reduced impact parameter $b_0$ modifies the relative magnitudes of the two contributions rather than merely rescaling their sum. Reducing the transverse smoothing width $\sigma_\perp$ generally strengthens both contributions, whereas increasing the initial energy-density normalization $e_{\mathrm{peak}}$ can preferentially weaken $P_{z,s2}^{\Xi}$ and reverse the total harmonic. At fixed $e_{\mathrm{peak}}$, the impact-parameter dependence is mainly shear driven for narrower transverse smoothing and mainly vorticity driven for broader transverse smoothing. Applying the Liu--Yin prescription to the two tested hydrodynamic backgrounds reverses the total sign, showing that the sign depends on the polarization formula as well as on the fluid fields. Our results obtained with ideal hydrodynamics and the ILE prescription have the same sign and order of magnitude as the ALICE measurements for peripheral Pb--Pb collisions in the $50$--$70\%$ centrality interval, showing that such a positive harmonic can arise from an initial-state-controlled, shear-dominated balance at decoupling.
\end{abstract}
\section{Introduction}

Relativistic hydrodynamics relates measured hadron spectra and anisotropic flow to the collective expansion of the matter produced in heavy-ion collisions \cite{ref11,ref12}. The velocity and temperature gradients associated with this expansion also enter the mean spin vector at leading order in local equilibrium \cite{ref13,ref14,ref15,ref42,ref20,ref21,ref61}. Hyperon polarization can therefore probe properties of the fluid that are not fixed by spectra and flow alone \cite{ref18,ref24,ref46}. The self-analyzing weak decay of the $\Lambda$ makes this polarization experimentally accessible. The global polarization measured by the STAR Collaboration established the connection between hyperon spin and the rotation of matter in noncentral collisions \cite{ref16,ref17,ref19,ref22,ref37}. By contrast, momentum-dependent, or local, polarization retains spatially varying information that is removed by the global momentum average \cite{ref43,ref45,ref48,ref49,ref36,ref54,ref55,ref56}.

At RHIC and LHC energies, anisotropic transverse expansion produces a quadrupolar longitudinal polarization along the beam direction. The corresponding second sine harmonic is

\[
P_{z,s2}=\langle P_z\sin[2(\phi-\Psi_2)]\rangle,
\]

where $\Psi_2$ is the second-order event-plane angle and the brackets denote a yield-weighted average within the experimental acceptance. Thermal-vorticity calculations originally predicted the opposite sign to the measured harmonic \cite{ref23,ref25}. The positive harmonic later measured by ALICE in Pb--Pb collisions at the LHC confirms that the local-polarization sign problem is not restricted to RHIC energies \cite{ref1}. The recent ALICE measurement at $\sqrt{s_{\mathrm{NN}}}=5.36\,\mathrm{TeV}$ provides the direct experimental reference for the present calculation \cite{ref75}.

The sign problem directs attention to the first-order gradients entering the polarization formula \cite{ref28,ref29,ref63}. Thermal vorticity is the antisymmetric derivative of the inverse-temperature four-vector $u^\mu/T$. By contrast, the symmetric derivative, known as thermal shear, generates an additional shear-induced polarization \cite{ref32}. Becattini, Buzzegoli, and Palermo (BBP) derived this contribution from local equilibrium and obtained an isothermal formulation for emission at fixed temperature \cite{ref2,ref3}. Liu and Yin derived a related first-order expression from Wigner functions and linear response \cite{ref4}. In numerical calculations, the shear and vorticity contributions can have opposite signs, and their competition can produce the observed positive harmonic \cite{ref27,ref40,ref44,ref47}. The resulting sign is set by their relative magnitudes, which depend on the velocity and temperature gradients at particle emission.

These emission-time gradients contain the combined effects of the initial energy-density distribution, hydrodynamic expansion, and decoupling condition. Previous calculations have connected local polarization to initial hot-spot size, anisotropic flow, viscous evolution, and cancellations between vorticity and shear \cite{ref38,ref52,ref35,ref74}. However, it remains unclear how changes in the transverse initial state are transmitted separately to the kinematic-vorticity contribution $P_{z,s2}^{\omega}$ and the kinematic-shear contribution $P_{z,s2}^{\Xi}$ within a single hydrodynamic framework.

We address this problem with the Particle-In-Cell Relativistic (PICR) ideal-hydrodynamic model and the BBP isothermal local-equilibrium (ILE) prescription. Twelve selected initial states vary the initial energy-density normalization $e_{\mathrm{peak}}$, reduced impact parameter $b_0$, and transverse smoothing width $\sigma_\perp$ and resolve the vorticity--shear cancellation boundary. We separate $P_{z,s2}^{\omega}$ and $P_{z,s2}^{\Xi}$ to determine how each initial-state parameter changes their competition, and we evaluate the Liu--Yin prescription on identical fluid fields to isolate polarization-formula dependence. We evaluate spin at emission without evolving an independent spin density or examining pseudogauge dependence \cite{ref30,ref31,ref39,ref62}; the hyperon sample contains directly emitted thermal hyperons and excludes resonance-decay feed-down \cite{ref70,ref71}.

The remainder of this paper is organized as follows. Section 2 introduces the BBP isothermal local-equilibrium and Liu--Yin polarization prescriptions. Section 3 describes the initial-state construction, PICR ideal-hydrodynamic evolution, and decoupling procedure. Section 4 defines the longitudinal-polarization and anisotropic-flow observables. Section 5 presents the calculated polarization, its comparison with the ALICE measurements, the initial-state dependence of the vorticity--shear competition, and the dependence on the polarization prescription. Section 6 tests robustness to the decoupling temperature and numerical reconstruction, examines sensitivity to pre-equilibrium transverse flow, and discusses the relation between kinematic shear and viscosity. Section 7 summarizes the main conclusions.

\section{Spin polarization at local equilibrium}

\subsection{Thermal vorticity and thermal shear}

We use the BBP isothermal local-equilibrium expression for the principal calculations and the Liu--Yin expression for comparison. In natural units, with $g^{\mu\nu}=\operatorname{diag}(1,-1,-1,-1)$ and $\epsilon^{0123}=+1$, we define the inverse-temperature four-vector as $\beta^\mu=u^\mu/T$, where $u^\mu$ is the fluid four-velocity and $T$ is the local temperature. Its first-order derivative separates into the antisymmetric thermal-vorticity tensor and the symmetric thermal-shear tensor,

\begin{equation}
\begin{aligned}
\varpi_{\mu\nu}&=\tfrac12(\partial_\nu\beta_\mu-\partial_\mu\beta_\nu),\\
\xi_{\mu\nu}&=\tfrac12(\partial_\mu\beta_\nu+\partial_\nu\beta_\mu).
\end{aligned}
\label{eq:thermal}
\end{equation}

The measured polarization receives contributions from particles emitted at different points of the decoupling hypersurface. To write the decoupling-hypersurface integrals for the spin contributions without repeating their statistical weights, we define the momentum-dependent surface average

\begin{equation}
\langle A\rangle_\Sigma\equiv
\frac{\int_\Sigma d\Sigma\!\cdot\!p\,n_F(1-n_F)A}
{\int_\Sigma d\Sigma\!\cdot\!p\,n_F},
\label{eq:average}
\end{equation}

where $n_F=[\exp(\beta\!\cdot\!p-\mu/T)+1]^{-1}$ is the Fermi--Dirac distribution. We set the chemical potential to zero and use the Boltzmann approximation, $n_F(1-n_F)\simeq n_F\simeq\exp(-p\!\cdot\!u/T)$. With these assumptions, the thermal-vorticity contribution to the mean spin four-vector of a spin-$1/2$ particle of mass $m$ is \cite{ref10}

\begin{equation}
S^\mu_\varpi(p)=-\frac{1}{8m}\epsilon^{\mu\rho\sigma\tau}p_\tau
\langle\varpi_{\rho\sigma}\rangle_\Sigma.
\label{eq:thermalspin}
\end{equation}

The symmetric tensor $\xi_{\mu\nu}$ produces an additional local-equilibrium contribution \cite{ref2}. This contribution is distinct from a viscous correction to the distribution function and remains present in ideal hydrodynamics. For a constant-temperature decoupling hypersurface, we use the isothermal form below.

\subsection{Isothermal local equilibrium}

Our calculation evaluates polarization on a hypersurface of fixed temperature, $T=T_{\mathrm{dec}}$. For this choice, the BBP isothermal expansion expresses the spin vector directly in terms of velocity gradients \cite{ref3}. Following Ref. \cite{ref52}, we define the antisymmetric and symmetric velocity-gradient tensors as kinematic vorticity and kinematic shear, respectively:

\begin{equation}
\begin{aligned}
\omega_{\rho\sigma}&=\tfrac12(\partial_\sigma u_\rho-\partial_\rho u_\sigma),\\
\Xi_{\rho\sigma}&=\tfrac12(\partial_\sigma u_\rho+\partial_\rho u_\sigma).
\end{aligned}
\label{eq:kinematic}
\end{equation}

The kinematic-vorticity tensor $\omega_{\rho\sigma}$ and the kinematic-shear tensor $\Xi_{\rho\sigma}$ generate separate contributions to the mean spin vector. In terms of the surface average in Eq.~\eqref{eq:average}, their sum in the ILE prescription is

\begin{equation}
S^\mu_{\mathrm{ILE}}(p)=-\frac{\epsilon^{\mu\rho\sigma\tau}p_\tau}{8mT_{\mathrm{dec}}}
\left[\langle\omega_{\rho\sigma}\rangle_\Sigma
+2\hat t_\rho\frac{p^\lambda}{p^0}\langle\Xi_{\lambda\sigma}\rangle_\Sigma\right].
\label{eq:ile}
\end{equation}

The vector $\hat t^\mu=(1,0,0,0)$ is the unit timelike vector defining the time direction in the collision center-of-mass frame, and $p^0=p\!\cdot\!\hat t$ is the particle energy in that frame. This reference vector is distinct from the local fluid four-velocity $u^\mu$, so the energy factor in Eq.~\eqref{eq:ile} is $p^0$ rather than $p\!\cdot\!u$. Superscripts $\omega$ and $\Xi$ label the kinematic-vorticity and kinematic-shear contributions to polarization. The tensor $\Xi_{\mu\nu}$ denotes the full symmetric derivative in Eq.~\eqref{eq:kinematic}; it differs from both the projected traceless shear tensor and the dissipative shear-stress tensor.

The absence of explicit temperature-gradient terms in Eq.~\eqref{eq:ile} follows from expanding the local-equilibrium density operator on an isothermal hypersurface \cite{ref3}. Since $T=T_{\mathrm{dec}}$ on $\Sigma$, the factor $1/T_{\mathrm{dec}}$ is taken outside the surface integral before expanding the four-velocity. The isothermal condition sets the tangential temperature derivatives to zero on $\Sigma$ while leaving the normal derivative unrestricted. The velocity derivatives entering Eq.~\eqref{eq:ile} are the full spacetime derivatives rather than projections onto the hypersurface. The observable polarization is obtained by boosting the mean spin vector to the hyperon rest frame and using $\boldsymbol P=2\boldsymbol S^{*}$ for the spin-$1/2$ $\Lambda$.

\subsection{The Liu--Yin polarization prescription}

The polarization obtained from a given hydrodynamic evolution also depends on the local-equilibrium formula used to evaluate the spin vector. We therefore calculate the polarization with the Liu--Yin (LY) formula on the same fluid fields and decoupling hypersurfaces used for the ILE calculation. The LY formula contains vorticity, temperature-gradient, chemical-potential-gradient, and shear contributions. Using the neutral ideal-fluid equations, its vorticity and temperature-gradient terms can be combined into thermal vorticity, as in Eq. (59) of Ref. \cite{ref4}. After neglecting gradients of $\mu/T$, the LY expression used in this paper is

\begin{equation}
S^\mu_{\mathrm{LY}}=S^\mu_\varpi+S^\mu_{\mathrm{LY,SIP}}.
\label{eq:ly}
\end{equation}

The thermal-vorticity term $S^\mu_\varpi$ combines the vorticity and temperature-gradient contributions of the original Liu--Yin formulation. The notation $S^\mu_{\mathrm{LY,SIP}}$ denotes the Liu--Yin shear-induced contribution, which contains the projected traceless tensor

\begin{equation}
\begin{aligned}
\sigma_{\mu\nu}&=\tfrac12(\nabla_\mu u_\nu+\nabla_\nu u_\mu)
-\tfrac13\Delta_{\mu\nu}\theta,\\
\Delta_{\mu\nu}&=g_{\mu\nu}-u_\mu u_\nu,\qquad
\nabla_\mu=\Delta_\mu{}^\alpha\partial_\alpha,\\
\theta&=\partial_\alpha u^\alpha,
\end{aligned}
\label{eq:lytensor}
\end{equation}

and an energy denominator proportional to $p\!\cdot\!u$, the particle energy in the local fluid rest frame \cite{ref4}. Unlike Eq.~\eqref{eq:ile}, the LY expression retains the temperature-gradient contribution through $S^\mu_\varpi$. To isolate the dependence on the polarization prescription, we evaluate the ILE and LY expressions on identical hydrodynamic fields and decoupling hypersurfaces. Because combining the original LY vorticity and temperature-gradient terms into $S^\mu_\varpi$ relies on continuum ideal-fluid relations, discretizing the combined and uncombined forms separately can yield different numerical integrals.

\begin{table*}[!t]
\centering
\caption{Model parameters, their physical roles, and the values used in the twelve selected initial states. The reference state uses $b_0=0.77$, $e_{\mathrm{peak}}=10\,\mathrm{GeV}/\mathrm{fm}^3$, and $\sigma_\perp=1.5\,\mathrm{fm}$.}\label{tab:parameters}
\setlength{\tabcolsep}{5pt}
\begin{tabular}{@{}lp{0.18\textwidth}p{0.25\textwidth}p{0.34\textwidth}@{}}
\toprule
Parameter & Definition & Role & Values \\
\midrule
$b_0$ & $b/(R_P+R_T)$ & Transverse geometry and initial eccentricity & selected scan: $0.60$, $0.77$ \\
$e_{\mathrm{peak}}$ & Pre-smoothing energy-density scale & Pressure scale and evolution lifetime & selected scan: $10$, $26$, $30$, $32$, $34$, $36$, $38$, $40\,\mathrm{GeV}/\mathrm{fm}^3$ \\
$\sigma_\perp$ & Gaussian width & Profile smoothing and local gradient strength & selected scan: $0.6$, $1.0$, $1.2$, $1.5\,\mathrm{fm}$ \\
$t_0$ & Cartesian matching time & Start of hydrodynamic evolution & $4\,\mathrm{fm}/c$ \\
$\eta_{\mathrm{flat}}$, $\sigma_\eta$ & Plateau and tail widths & Longitudinal profile shape & $0.8$, $0.6$ \\
$T_{\mathrm{dec}}$ & Freeze-out temperature & Isothermal emission surface & reference: $160\,\mathrm{MeV}$; scan: $157.5$, $158.5$, $170\,\mathrm{MeV}$ \\
\bottomrule
\end{tabular}
\end{table*}

\section{Initial state and hydrodynamic evolution}

\subsection{Transverse geometry and smoothing}

To study how the transverse initial energy density affects subsequent polarization, we vary the collision geometry and its transverse smoothing width. The Magas--Csernai--Strottman (MCS) initial-state model represents the Lorentz-contracted projectile and target nuclei by transverse cells \cite{ref6,ref7,ref8,ref9}; $W_{\mathrm{MCS}}(x,y)$ is the unsmoothed energy weight of the participating matter in the cell at transverse position $(x,y)$. We obtain a continuous transverse profile by a two-dimensional Gaussian convolution:

\begin{equation}
\begin{aligned}
W_\perp(x,y)={}&\mathcal N\int dx'dy'\,W_{\mathrm{MCS}}(x',y')\\
&\times\exp\left[-\frac{(x-x')^2+(y-y')^2}{2\sigma_\perp^2}\right].
\end{aligned}
\label{eq:smoothing}
\end{equation}

The unsmoothed MCS weight is normalized to its maximum, and $\mathcal N$ is chosen so that the Gaussian convolution preserves the transverse integral. Thus, at fixed $b_0$ and $e_{\mathrm{peak}}$, changing $\sigma_\perp$ redistributes the transverse energy density without changing its integral. A larger $\sigma_\perp$ produces a broader profile and weaker local spatial variations. We therefore use $\sigma_\perp$ as an effective smoothing scale rather than an individual-nucleon radius.

To quantify how much geometric anisotropy remains after smoothing the initial energy-density distribution, we use the energy-weighted initial eccentricity $\epsilon_{2,\mathrm{init}}=\langle y^2-x^2\rangle_W/\langle y^2+x^2\rangle_W$. The $x$ axis lies along the impact parameter and the $y$ axis is perpendicular to the reaction plane. We compare two reduced impact parameters, $b_0=b/(R_P+R_T)=0.60$ and $0.77$, where $R_P$ and $R_T$ are the projectile and target nuclear radii. For orientation, the geometrical estimate $c\simeq b_0^2$ associates $b_0=0.77$ with $c\simeq59\%$; all calculations use fixed $b_0$ rather than centrality-averaged events. Evaluating the two $b_0$ values at the same transverse smoothing width $\sigma_\perp$ separates the effect of changing $b_0$ from that of changing $\sigma_\perp$.

\Needspace{8\baselineskip}
\subsection{Longitudinal profile and initialization}

The transverse energy density profile must be supplemented by a longitudinal profile and an initial velocity field to define the three-dimensional hydrodynamic initial condition. We combine the transverse weight with a longitudinal envelope according to

\begin{equation}
e(x,y,\eta_s)=e_{\mathrm{peak}}W_\perp(x,y)H(\eta_s),
\label{eq:initial}
\end{equation}

where $\eta_s=\operatorname{atanh}(z/t_0)$ is the spacetime rapidity on the initial Cartesian-time slice and $H(\eta_s)$ consists of a central plateau and Gaussian tails. The initial energy-density normalization $e_{\mathrm{peak}}$ sets the scale of the profile before transverse smoothing with width $\sigma_\perp$. Consequently, $e_{\mathrm{peak}}$ need not equal the maximum energy density after initialization, which also depends on $\sigma_\perp$ and $b_0$. The scan tests polarization sensitivity rather than fitting the measured multiplicity or transverse energy.

The initial longitudinal flow rapidity is set to the Bjorken value, $y_f=\eta_s$ \cite{ref72}, and the initial transverse velocity is set to zero. Only the transverse energy-density distribution is therefore retained from the MCS model. The longitudinal stopping dynamics of the original MCS construction is replaced by the prescribed Bjorken flow, with vanishing tilt and rapidity-shift parameters. The resulting initial condition is reflection symmetric and carries no net angular momentum, $J_y=0$, which isolates the effect of the transverse energy-density geometry from polarization produced by a tilted longitudinal source. The Cartesian time $t_0=4\,\mathrm{fm}/c$ labels the slice supplied to PICR, and the zero initial transverse velocity omits transverse expansion before this slice. Table~\ref{tab:parameters} summarizes the model parameters.

\begin{figure*}[!t]
\centering
\includegraphics[width=0.65\textwidth]{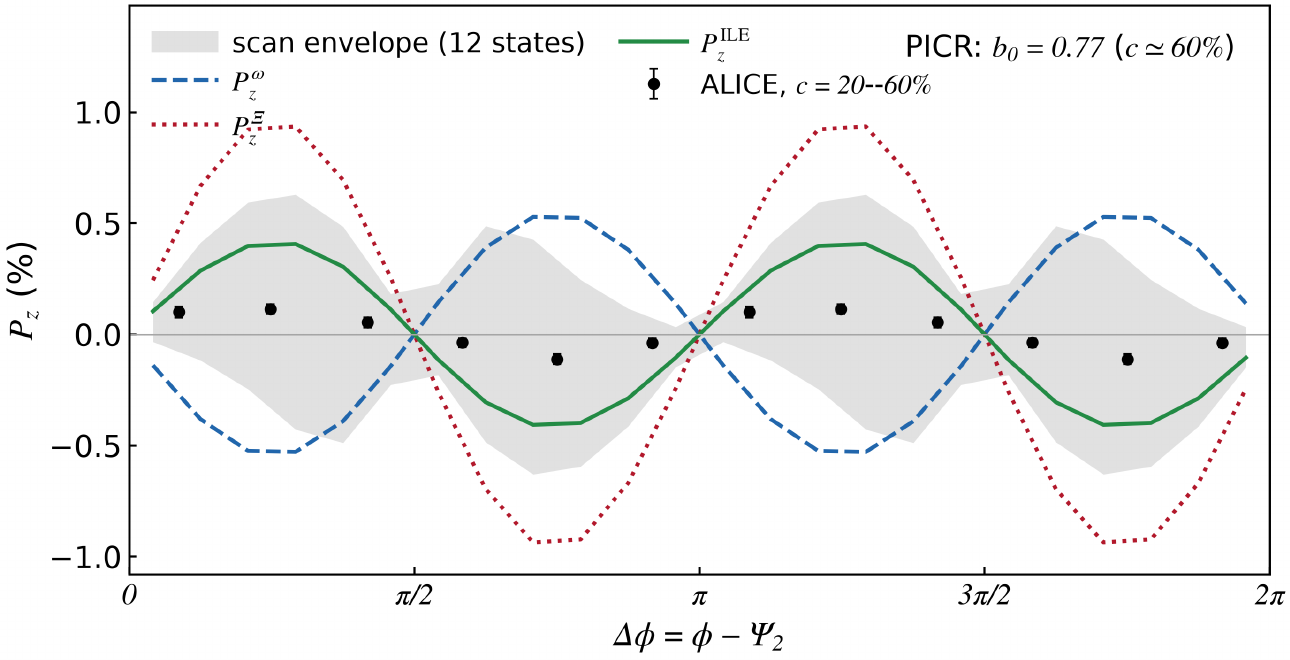}
\caption{Azimuthal longitudinal polarization for the reference state $b_0=0.77$, $e_{\mathrm{peak}}=10\,\mathrm{GeV}/\mathrm{fm}^3$, and $\sigma_\perp=1.5\,\mathrm{fm}$, where $\Delta\phi=\phi-\Psi_2$. The curves show the kinematic-vorticity contribution $P_z^{\omega}$, kinematic-shear contribution $P_z^{\Xi}$, and total $P_z^{\mathrm{ILE}}$. The gray band is the envelope of the total ILE polarization curves for the twelve initial states listed in Table~\ref{tab:scan}. The calculated curves are integrated over $|y|<0.5$ and $0.5<p_T<6\,\mathrm{GeV}/c$. The black points are approximately digitized ALICE data for $20$--$60\%$ Pb--Pb collisions at $5.36\,\mathrm{TeV}$ \cite{ref75}. They are converted from the published variable $2(\phi-\Psi_2)$ and repeated over the second period only for visual comparison.}
\label{fig:phi}
\end{figure*}

\subsection{Ideal evolution and freeze-out}

Starting from the energy density and velocity fields defined above, PICR evolves the system until it reaches the chosen decoupling temperature. The evolution is governed by energy-momentum conservation for an ideal fluid,

\begin{equation}
\begin{aligned}
\partial_\mu T^{\mu\nu}&=0,\\
T^{\mu\nu}&=(e+p)u^\mu u^\nu-pg^{\mu\nu}.
\end{aligned}
\label{eq:hydro}
\end{equation}

The evolution contains no explicit shear- or bulk-viscous stress, although the numerical scheme may introduce dissipation. To close the ideal-hydrodynamic equations, we use the three-flavor bag-model equation of state implemented in PICR \cite{ref73}. At $T_{\mathrm{dec}}=160\,\mathrm{MeV}$, it gives $p/e=0.027$, compared with $p/e\simeq0.17$--$0.20$ from lattice QCD in the same temperature region \cite{ref76}. The bag-model equation of state does not reproduce the QCD crossover, which limits quantitative comparison with data. The reported polarization therefore corresponds to the pressure gradients and decoupling surface generated by this equation of state.

To calculate the emitted-particle spectrum and the polarization integrals, we identify the decoupling hypersurface by the condition $T=T_{\mathrm{dec}}$. Its oriented surface elements are reconstructed in four-dimensional spacetime with Cornelius++ \cite{ref5}. The hyperon spectrum, which also supplies the yield weights in the polarization averages, is obtained from the Cooper--Frye integral \cite{ref34},

\begin{equation}
E\frac{dN}{d^3p}=\frac{g}{(2\pi)^3}\int_\Sigma p^\mu d\Sigma_\mu f_0(x,p).
\label{eq:cf}
\end{equation}

In the Cooper--Frye expression, $E=p^0$ is the particle energy, $g$ is the spin degeneracy, and $f_0(x,p)$ is the local-equilibrium distribution evaluated with the Boltzmann approximation specified above. The factor $p^\mu d\Sigma_\mu$ is retained with its sign in both the particle-yield and polarization integrals, including surface elements for which it is negative. Restricting the integral to positive flux would define a different emission prescription. Such integrals are used only for numerical diagnostics.

The spin calculation also requires velocity gradients at each surface element. The gradients and the Cornelius++ surface geometry are obtained by separate reconstruction procedures. The numerical effects of their relative placement and the residual nonclosure of the reconstructed hypersurface are examined in Sec.~\ref{sec:discussion}.

\section{Observables}

\label{sec:observables}

To facilitate comparison with LHC measurements, we specify the momentum acceptance and averaging procedure used to construct the observables. Unless stated otherwise, the integrated results throughout this paper refer to momentum rapidity $|y|<0.5$ and transverse momentum $0.5<p_T<6\,\mathrm{GeV}/c$. The transverse-momentum integration measure is defined as

\begin{equation}
d\Gamma=p_Tdp_Td\phi\,F(p_T,\phi).
\label{eq:measure}
\end{equation}

The function $F(p_T,\phi)$ is the invariant hyperon yield averaged over the accepted rapidity interval. This invariant yield weights each momentum bin according to its hyperon yield.

Using this measure, the elliptic-flow coefficient $v_2$ and the polarization sine harmonics $P_{z,sn}$ are defined as

\begin{equation}
v_2=\frac{\int d\Gamma\,\cos[2(\phi-\Psi_2)]}{\int d\Gamma},
\label{eq:v2}
\end{equation}

\begin{equation}
P_{z,sn}=\frac{\int d\Gamma\,P_z(p_T,\phi)\sin[n(\phi-\Psi_2)]}{\int d\Gamma},
\label{eq:harmonic}
\end{equation}

where $P_z(p_T,\phi)$ denotes the longitudinal polarization averaged over the accepted rapidity interval with the hyperon yield as weight, and $\Psi_2$ is the second-order event-plane angle. Momentum rapidity $y$ in these definitions is distinct from the transverse coordinate used in the initial energy density profile. The two observables share the same yield weight, but $v_2$ measures the azimuthal anisotropy of the particle distribution, whereas $P_{z,sn}$ measures the corresponding sine moment of its longitudinal polarization.

The weights in Eqs.~\eqref{eq:v2} and \eqref{eq:harmonic} are obtained from our calculated hyperon spectrum with signed Cooper--Frye flux, unless otherwise specified. To describe how the kinematic-vorticity and kinematic-shear contributions combine, we write $P_{z,s2}^{\mathrm{ILE}}=P_{z,s2}^{\omega}+P_{z,s2}^{\Xi}$ and define $R=\lvert P_{z,s2}^{\Xi}/P_{z,s2}^{\omega}\rvert$. The ratio $R$ compares the magnitude of the kinematic-shear contribution $P_{z,s2}^{\Xi}$ with the magnitude of the kinematic-vorticity contribution $P_{z,s2}^{\omega}$. All initial states examined below have $P_{z,s2}^{\omega}<0$ and $P_{z,s2}^{\Xi}>0$. Under these conditions, $R>1$ gives a positive total ILE harmonic, $R<1$ gives a negative total ILE harmonic, and $R=1$ gives exact cancellation.

\begin{figure}[!htbp]
\centering
\includegraphics[width=\columnwidth]{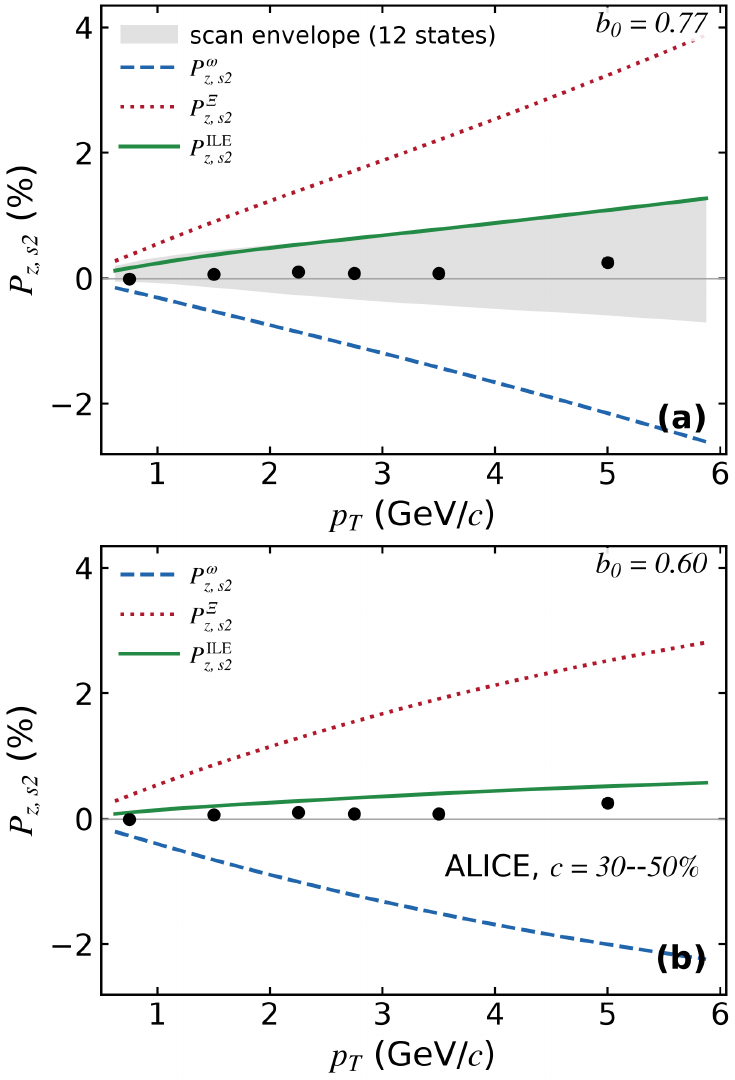}
\caption{Transverse-momentum dependence for (a) $b_0=0.77$ and (b) $b_0=0.60$. The three calculated curves in each panel show $P_{z,s2}^{\omega}(p_T)$, $P_{z,s2}^{\Xi}(p_T)$, and $P_{z,s2}^{\mathrm{ILE}}(p_T)$. They use $e_{\mathrm{peak}}=10\,\mathrm{GeV}/\mathrm{fm}^3$, $\sigma_\perp=1.5\,\mathrm{fm}$, and $|y|<0.5$. The gray band in panel (a) is the envelope of the total ILE polarization curves for the twelve initial states listed in Table~\ref{tab:scan}. The black points are approximately digitized ALICE data for $30$--$50\%$ Pb--Pb collisions at $5.36\,\mathrm{TeV}$ \cite{ref75}. Neither the calculation nor the data are rescaled. Our results use the calculated spectrum of directly emitted $\Lambda+\bar{\Lambda}$ and signed Cooper--Frye flux.}
\label{fig:decomp}
\end{figure}

\section{Results}

\subsection{A positive quadrupole from competing contributions}

\label{sec:reference_result}

We first establish the polarization mechanism in a reference calculation with $b_0=0.77$, $e_{\mathrm{peak}}=10\,\mathrm{GeV}/\mathrm{fm}^3$, and $\sigma_\perp=1.5\,\mathrm{fm}$. At $T_{\mathrm{dec}}=160\,\mathrm{MeV}$, the ILE contributions integrated over the acceptance in Sec.~\ref{sec:observables} are
\begin{equation}
\begin{aligned}
P_{z,s2}^{\omega}&=-2.72\times10^{-3},\\
P_{z,s2}^{\Xi}&=+4.80\times10^{-3},\\
P_{z,s2}^{\mathrm{ILE}}&=+2.08\times10^{-3}.
\end{aligned}
\label{eq:reference}
\end{equation}

The negative kinematic-vorticity contribution $P_{z,s2}^{\omega}$ cancels more than half of the positive kinematic-shear contribution $P_{z,s2}^{\Xi}$. Because the competition ratio $R=\lvert P_{z,s2}^{\Xi}/P_{z,s2}^{\omega}\rvert=1.77$, kinematic shear remains larger and the total ILE harmonic $P_{z,s2}^{\mathrm{ILE}}$ is positive. This vorticity--shear competition is consistent with earlier ILE studies \cite{ref3,ref52}. Since the sign is controlled by the relative magnitudes of the two contributions, the initial-state variations studied below can reverse the total harmonic by changing the two freeze-out hypersurface integrals differently.

Figure~\ref{fig:phi} compares the calculated azimuthal dependence with the ALICE data at $5.36\,\mathrm{TeV}$ \cite{ref75}. The reference value $b_0=0.77$ corresponds approximately to $c\simeq59\%$, and the calculated positive harmonic $P_{z,s2}^{\mathrm{ILE}}=2.08\times10^{-3}$ has the same sign and order of magnitude as the ALICE measurements for peripheral collisions at $c=50$--$70\%$. This agreement indicates that a shear-dominated balance at isothermal local equilibrium can account for the positive longitudinal-polarization signal observed in peripheral collisions.

Figure~\ref{fig:decomp} extends the comparison to $P_{z,s2}(p_T)$. At both $b_0=0.77$ and $0.60$, the positive kinematic-shear contribution remains larger than the negative kinematic-vorticity contribution. The calculated total harmonics and the ALICE measurement for Pb--Pb collisions at $5.36\,\mathrm{TeV}$ and $c=30$--$50\%$ have the same positive sign, increase overall with $p_T$, and have the same order of magnitude \cite{ref75}. The smaller calculated harmonics for $b_0=0.60$, a value corresponding approximately to $c\simeq36\%$, are also consistent with the experimental increase toward more peripheral collisions. Again, these agreements indicate that the shear-dominated ILE response captures the sign, scale, and overall $p_T$ dependence of the measured longitudinal polarization.

\subsection{Initial-state control of the vorticity--shear balance}

\label{sec:initial_state_scan}

\begin{figure}[!htbp]
\centering
\includegraphics[width=\columnwidth]{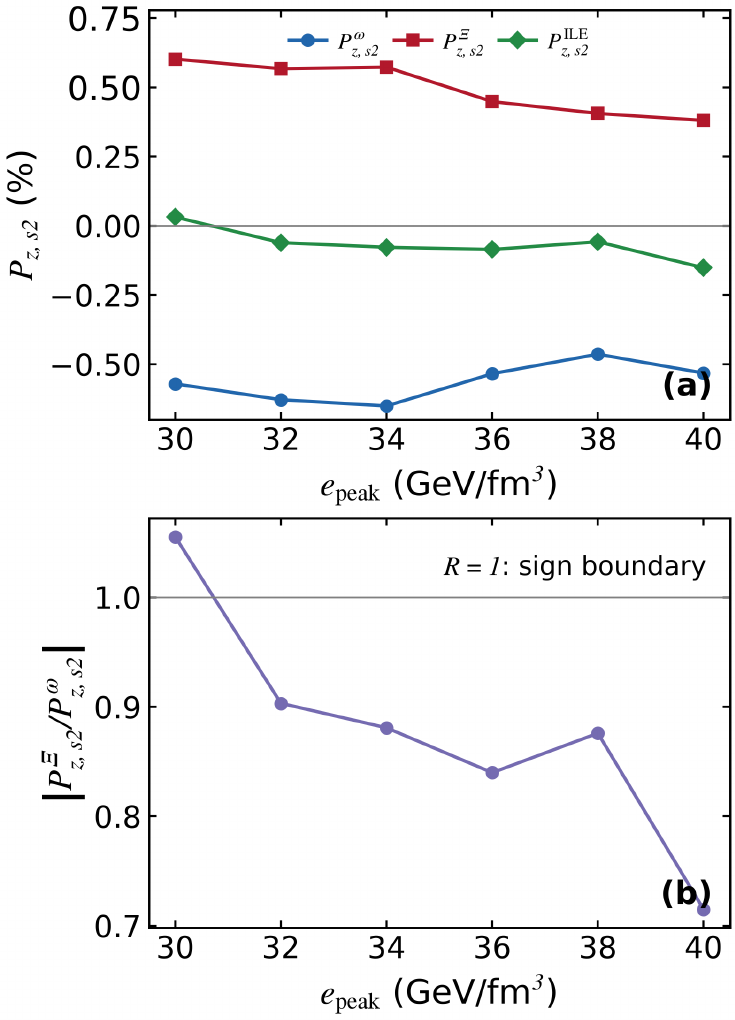}
\caption{Dependence on the initial energy-density normalization $e_{\mathrm{peak}}$ at fixed $b_0=0.77$ and $\sigma_\perp=0.6\,\mathrm{fm}$. Panel (a) shows the momentum-integrated kinematic-vorticity contribution $P_{z,s2}^{\omega}$, kinematic-shear contribution $P_{z,s2}^{\Xi}$, and total ILE harmonic $P_{z,s2}^{\mathrm{ILE}}$. Panel (b) shows $R=\lvert P_{z,s2}^{\Xi}/P_{z,s2}^{\omega}\rvert$. The polarization contributions are integrated over $|y|<0.5$ and $0.5<p_T<6\,\mathrm{GeV}/c$. Horizontal lines mark zero polarization and the cancellation boundary $R=1$. Lines connect our calculated points only to guide the eye.}
\label{fig:scan}
\end{figure}

The reference calculation in Sec.~\ref{sec:reference_result} shows how a positive total harmonic results from the competition between the negative kinematic-vorticity contribution $P_{z,s2}^{\omega}$ and the positive kinematic-shear contribution $P_{z,s2}^{\Xi}$. To determine how the initial transverse energy density changes this competition, we vary the initial energy-density normalization $e_{\mathrm{peak}}$ and the transverse smoothing width $\sigma_\perp$. Table~\ref{tab:scan} and Fig.~\ref{fig:scan} summarize the results. The signs of the two contributions remain unchanged throughout the scan: $P_{z,s2}^{\omega}<0$ and $P_{z,s2}^{\Xi}>0$. The ratio $R=\lvert P_{z,s2}^{\Xi}/P_{z,s2}^{\omega}\rvert$ therefore determines the sign of the total ILE harmonic, which is positive for $R>1$ and negative for $R<1$. The scan tests whether $e_{\mathrm{peak}}$ and $\sigma_\perp$ change the relative magnitudes enough to move the total harmonic across the cancellation boundary $R=1$.

\begin{table*}[!t]
\centering
\caption{Comparison of the twelve initial states used to construct the gray envelopes in Fig.~\ref{fig:phi} and Fig.~\ref{fig:decomp}(a), ordered by the unrounded elliptic flow $v_2$. The initial energy-density normalization $e_{\mathrm{peak}}$ is in $\mathrm{GeV}/\mathrm{fm}^3$ and the transverse smoothing width $\sigma_\perp$ in $\mathrm{fm}$. The elliptic flow $v_2$ and polarization entries are evaluated over $|y|<0.5$ and $0.5<p_T<6\,\mathrm{GeV}/c$. Polarization entries are in units of $10^{-3}$ and are evaluated with our calculated spectrum and signed Cooper--Frye flux. The ratio is $R=\lvert P_{z,s2}^{\Xi}/P_{z,s2}^{\omega}\rvert$. Totals are computed from unrounded component values and may differ in the last displayed digit from the sum of the rounded components.}\label{tab:scan}
\setlength{\tabcolsep}{5pt}
\begin{tabular}{@{}rrrrrrrr@{}}
\toprule
$b_0$ & $e_{\mathrm{peak}}$ & $\sigma_\perp$ & $v_2$ & $10^3P_{z,s2}^{\omega}$ & $10^3P_{z,s2}^{\Xi}$ & $10^3P_{z,s2}^{\mathrm{ILE}}$ & $R$ \\
\midrule
$0.77$ & $10$ & $1.5$ & $0.054$ & $-2.721$ & $+4.804$ & $+2.083$ & $1.77$ \\
$0.60$ & $10$ & $1.5$ & $0.086$ & $-3.764$ & $+5.045$ & $+1.281$ & $1.34$ \\
$0.77$ & $10$ & $1.0$ & $0.140$ & $-5.199$ & $+8.319$ & $+3.120$ & $1.60$ \\
$0.60$ & $10$ & $1.0$ & $0.140$ & $-5.033$ & $+6.494$ & $+1.461$ & $1.29$ \\
$0.77$ & $26$ & $1.2$ & $0.160$ & $-3.965$ & $+4.714$ & $+0.749$ & $1.19$ \\
$0.77$ & $30$ & $1.2$ & $0.164$ & $-4.005$ & $+4.158$ & $+0.154$ & $1.04$ \\
$0.77$ & $30$ & $0.6$ & $0.295$ & $-5.707$ & $+6.023$ & $+0.316$ & $1.06$ \\
$0.77$ & $36$ & $0.6$ & $0.300$ & $-5.342$ & $+4.487$ & $-0.855$ & $0.84$ \\
$0.77$ & $32$ & $0.6$ & $0.301$ & $-6.281$ & $+5.674$ & $-0.608$ & $0.90$ \\
$0.77$ & $34$ & $0.6$ & $0.305$ & $-6.506$ & $+5.731$ & $-0.775$ & $0.88$ \\
$0.77$ & $38$ & $0.6$ & $0.306$ & $-4.635$ & $+4.060$ & $-0.575$ & $0.88$ \\
$0.77$ & $40$ & $0.6$ & $0.308$ & $-5.320$ & $+3.804$ & $-1.516$ & $0.72$ \\
\bottomrule
\end{tabular}
\end{table*}

Table~\ref{tab:scan} shows that, at fixed $e_{\mathrm{peak}}=10\,\mathrm{GeV}/\mathrm{fm}^3$, reducing the transverse smoothing width $\sigma_\perp$ from $1.5$ to $1.0\,\mathrm{fm}$ strengthens both polarization contributions at both impact parameters. The positive kinematic-shear contribution increases by a larger absolute amount than the magnitude of the negative kinematic-vorticity contribution, so the total ILE harmonic increases. The same strengthening occurs between $\sigma_\perp=1.2$ and $0.6\,\mathrm{fm}$ at $e_{\mathrm{peak}}=30\,\mathrm{GeV}/\mathrm{fm}^3$, although the competition ratio $R$ changes little because both contributions grow. The transverse smoothing width $\sigma_\perp$ therefore controls the magnitudes of the two polarization contributions, while their relative response determines the total sign. Furthermore, the last rows of Table~\ref{tab:scan} show that $R$ is already close to $1$ at $e_{\mathrm{peak}}=30\,\mathrm{GeV}/\mathrm{fm}^3$. Increasing $e_{\mathrm{peak}}$ from $30$ to $40\,\mathrm{GeV}/\mathrm{fm}^3$ drives $R$ across the cancellation boundary $R=1$ and changes the total ILE harmonic from positive to negative. Figure~\ref{fig:scan} displays this sign reversal more clearly.

Figure~\ref{fig:scan} shows that, at fixed $b_0=0.77$ and $\sigma_\perp=0.6\,\mathrm{fm}$, increasing $e_{\mathrm{peak}}$ from $30$ to $32\,\mathrm{GeV}/\mathrm{fm}^3$ lowers $R$ from $1.06$ to $0.90$ and reverses the total harmonic. All five points from $32$ to $40\,\mathrm{GeV}/\mathrm{fm}^3$ remain vorticity dominated despite nonmonotonic changes of the individual contributions. The component decomposition identifies the cause. Raising $e_{\mathrm{peak}}$ from $30$ to $40\,\mathrm{GeV}/\mathrm{fm}^3$ reduces the positive kinematic-shear contribution $P_{z,s2}^{\Xi}$ by $2.22\times10^{-3}$, while the negative kinematic-vorticity contribution $P_{z,s2}^{\omega}$ becomes $0.39\times10^{-3}$ less negative and partly offsets this decrease. A milder change from $26$ to $30\,\mathrm{GeV}/\mathrm{fm}^3$ at $\sigma_\perp=1.2\,\mathrm{fm}$ (see Table~\ref{tab:scan}) also shows a preferential reduction of the positive kinematic-shear contribution $P_{z,s2}^{\Xi}$ without reversing the total sign. Increasing the initial energy-density normalization $e_{\mathrm{peak}}$ can therefore reduce $P_{z,s2}^{\Xi}$ more strongly than $\lvert P_{z,s2}^{\omega}\rvert$ and thereby drive the total ILE harmonic $P_{z,s2}^{\mathrm{ILE}}$ across the cancellation boundary $R=1$, reversing its sign.

\subsection{Initial-state control of the vorticity--shear competition: initial eccentricity and elliptic flow}

\begin{figure}[!htbp]
\centering
\includegraphics[width=\columnwidth]{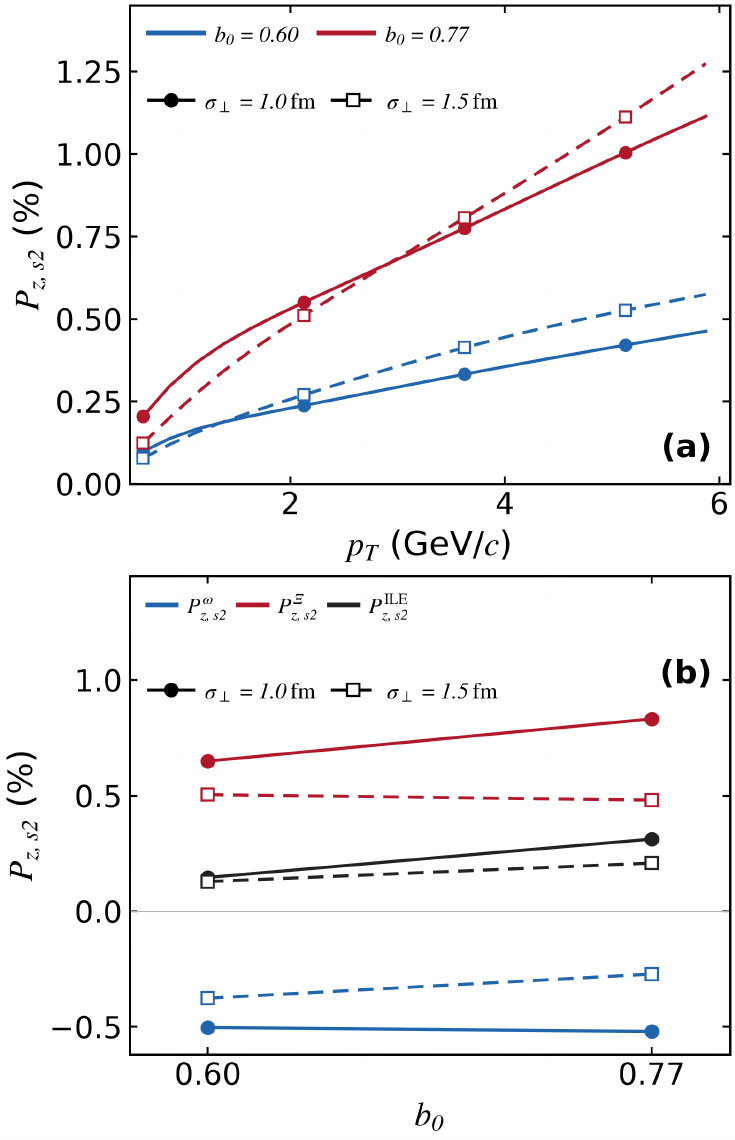}
\caption{Dependence on the reduced impact parameter $b_0$ and transverse smoothing width $\sigma_\perp$ at fixed $J_y=0$ and $e_{\mathrm{peak}}=10\,\mathrm{GeV}/\mathrm{fm}^3$. Panel (a) compares $P_{z,s2}^{\mathrm{ILE}}(p_T)$ at $b_0=0.60$ and $0.77$; colors distinguish the two impact parameters. Panel (b) separates the momentum-integrated kinematic-vorticity contribution $P_{z,s2}^{\omega}$, kinematic-shear contribution $P_{z,s2}^{\Xi}$, and total ILE harmonic $P_{z,s2}^{\mathrm{ILE}}$; colors distinguish the three polarization quantities. All calculated quantities use $|y|<0.5$; the momentum-integrated quantities in panel (b) also use $0.5<p_T<6\,\mathrm{GeV}/c$. In both panels, filled circles with solid lines denote $\sigma_\perp=1.0\,\mathrm{fm}$, whereas open squares with dashed lines denote $\sigma_\perp=1.5\,\mathrm{fm}$. Markers in panel (a) are shown only at selected momenta. Lines in panel (b) connect our results at the two impact parameters only to guide the eye.}
\label{fig:geometry}
\end{figure}

Table~\ref{tab:scan} also shows that the vorticity--shear balance is sensitive to the reduced impact parameter $b_0$. The reduced impact parameter $b_0$ changes the transverse shape of the initial energy-density profile. We quantify the resulting spatial anisotropy by the energy-weighted initial eccentricity $\epsilon_{2,\mathrm{init}}=\langle y^2-x^2\rangle_W/\langle y^2+x^2\rangle_W$, which measures the difference between the out-of-plane and in-plane transverse widths of the profile. During the hydrodynamic expansion, anisotropic pressure gradients convert part of this initial spatial anisotropy into momentum anisotropy, measured by the elliptic flow $v_2$. By varying $b_0$, we next compare the responses of the kinematic-vorticity contribution $P_{z,s2}^{\omega}$ and the kinematic-shear contribution $P_{z,s2}^{\Xi}$ with the initial eccentricity $\epsilon_{2,\mathrm{init}}$ and the elliptic flow $v_2$. This comparison tests whether the two polarization contributions can be connected to the starting spatial anisotropy or the momentum anisotropy developed during the expansion. The impact-parameter results are shown in Fig.~\ref{fig:geometry}, while Table~\ref{tab:energycontrast} subsequently tests whether this correspondence persists when $e_{\mathrm{peak}}$ changes. Since our simulations assume vanishing system angular momentum $J_y=0$, varying $b_0$ changes the transverse energy-density geometry without introducing longitudinal tilt.

The solid lines in Fig.~\ref{fig:geometry}(a) show that, at $\sigma_\perp=1.0\,\mathrm{fm}$, the total ILE harmonic $P_{z,s2}^{\mathrm{ILE}}$ is larger for $b_0=0.77$ than for $b_0=0.60$ throughout the plotted $p_T$ range. Figure~\ref{fig:geometry}(b) explains this difference. When $b_0$ is increased from $0.60$ to $0.77$, the negative kinematic-vorticity contribution $P_{z,s2}^{\omega}$ remains nearly unchanged, whereas the positive kinematic-shear contribution $P_{z,s2}^{\Xi}$ increases by about $30\%$. Because the two contributions partially cancel, the increase in $P_{z,s2}^{\Xi}$ makes the total ILE harmonic $P_{z,s2}^{\mathrm{ILE}}$ about twice as large. The integrated elliptic flow $v_2$, however, remains $0.14$ at both impact parameters, whereas the initial eccentricity $\epsilon_{2,\mathrm{init}}$ increases from $0.42$ to $0.48$. This comparison therefore suggests that, for $\sigma_\perp=1.0\,\mathrm{fm}$, the increase of $P_{z,s2}^{\Xi}$, and hence the increase of $P_{z,s2}^{\mathrm{ILE}}$, with $b_0$ is associated with the larger initial eccentricity $\epsilon_{2,\mathrm{init}}$ rather than with a change in the final elliptic flow $v_2$.

\begin{table*}[!t]
\centering
\caption{Comparison at fixed $b_0=0.77$ and transverse smoothing width $\sigma_\perp=1.2\,\mathrm{fm}$ used to test whether the initial eccentricity $\epsilon_{2,\mathrm{init}}$ or elliptic flow $v_2$ alone describes both polarization contributions when the initial energy-density normalization $e_{\mathrm{peak}}$ changes. The initial energy-density normalization $e_{\mathrm{peak}}$ is in $\mathrm{GeV}/\mathrm{fm}^3$. The elliptic flow $v_2$ and polarization entries are evaluated over $|y|<0.5$ and $0.5<p_T<6\,\mathrm{GeV}/c$. Polarization entries are in units of $10^{-3}$ and are evaluated with our calculated spectrum and signed Cooper--Frye flux. The ratio is $R=\lvert P_{z,s2}^{\Xi}/P_{z,s2}^{\omega}\rvert$. Totals are computed from unrounded component values and may differ in the last displayed digit from the sum of the rounded components.}\label{tab:energycontrast}
\setlength{\tabcolsep}{5pt}
\begin{tabular}{@{}rrrrrrr@{}}
\toprule
$e_{\mathrm{peak}}$ & $\epsilon_{2,\mathrm{init}}$ & $v_2$ & $10^3P_{z,s2}^{\omega}$ & $10^3P_{z,s2}^{\Xi}$ & $10^3P_{z,s2}^{\mathrm{ILE}}$ & $R$ \\
\midrule
$10$ & $0.414$ & $0.097$ & $-4.122$ & $+6.220$ & $+2.098$ & $1.51$ \\
$30$ & $0.414$ & $0.164$ & $-4.005$ & $+4.158$ & $+0.154$ & $1.04$ \\
\bottomrule
\end{tabular}
\end{table*}

Figure~\ref{fig:geometry}(b) also shows that the transverse smoothing width $\sigma_\perp$ changes both the strength and the origin of the impact-parameter dependence. Increasing $\sigma_\perp$ from $1.0$ to $1.5\,\mathrm{fm}$ reduces the increase of the total ILE harmonic $P_{z,s2}^{\mathrm{ILE}}$ from a factor of about $2$ to a factor of $1.6$, so the dependence on $b_0$ becomes weaker after broader transverse smoothing. At $\sigma_\perp=1.5\,\mathrm{fm}$, raising $b_0$ from $0.60$ to $0.77$ decreases the positive kinematic-shear contribution $P_{z,s2}^{\Xi}$ by only about $5\%$, whereas the magnitude of the negative kinematic-vorticity contribution $\lvert P_{z,s2}^{\omega}\rvert$ decreases by about $28\%$, as listed in Table~\ref{tab:scan}. The smaller magnitude of $P_{z,s2}^{\omega}$ therefore weakens its cancellation of $P_{z,s2}^{\Xi}$ and accounts for most of the remaining 1.6-fold increase in the total ILE harmonic $P_{z,s2}^{\mathrm{ILE}}$. At the same time, the initial eccentricity remains $\epsilon_{2,\mathrm{init}}\simeq0.33$ at both $b_0$ values, whereas the elliptic flow decreases from $v_2=0.086$ to $0.054$. This comparison therefore suggests that, at $\sigma_\perp=1.5\,\mathrm{fm}$, the $b_0$ dependence of $P_{z,s2}^{\mathrm{ILE}}$ is carried mainly by $P_{z,s2}^{\omega}$ and is associated with the change in $v_2$ rather than the nearly unchanged $\epsilon_{2,\mathrm{init}}$. The $\sigma_\perp=1.0$ and $1.5\,\mathrm{fm}$ comparisons together show that transverse smoothing can change the impact-parameter response from a shear-driven change to a vorticity-driven change.

The comparison in Fig.~\ref{fig:geometry} suggests that, at fixed initial energy-density normalization $e_{\mathrm{peak}}$, the kinematic-shear contribution $P_{z,s2}^{\Xi}$ is more closely connected to the initial spatial eccentricity $\epsilon_{2,\mathrm{init}}$, whereas the kinematic-vorticity contribution $P_{z,s2}^{\omega}$ is more closely associated with the momentum anisotropy $v_2$. However, Table~\ref{tab:energycontrast} shows that this simple correspondence does not persist when $e_{\mathrm{peak}}$ also changes. At fixed $b_0=0.77$, transverse smoothing width $\sigma_\perp=1.2\,\mathrm{fm}$, and initial eccentricity $\epsilon_{2,\mathrm{init}}$, increasing $e_{\mathrm{peak}}$ from $10$ to $30\,\mathrm{GeV}/\mathrm{fm}^3$ raises the elliptic flow $v_2$ by about $70\%$. Nevertheless, the magnitude of the kinematic-vorticity contribution $\lvert P_{z,s2}^{\omega}\rvert$ changes by only $3\%$, whereas the kinematic-shear contribution $P_{z,s2}^{\Xi}$ decreases by about $30\%$. The large change in $v_2$ without a comparable change in $\lvert P_{z,s2}^{\omega}\rvert$, together with the large change in $P_{z,s2}^{\Xi}$ at unchanged $\epsilon_{2,\mathrm{init}}$, shows that neither the final elliptic flow $v_2$ nor the initial eccentricity $\epsilon_{2,\mathrm{init}}$ alone describes the two polarization contributions across changes in $e_{\mathrm{peak}}$. Together with the ILE expression, this comparison indicates that the dependence on $e_{\mathrm{peak}}$ is transmitted through changes in the decoupling-time velocity-gradient fields and the decoupling hypersurface that are not encoded in either $\epsilon_{2,\mathrm{init}}$ or $v_2$ alone.

\begin{table*}[!t]
\centering
\caption{Polarization prescriptions evaluated at initial energy-density normalization $e_{\mathrm{peak}}=10\,\mathrm{GeV}/\mathrm{fm}^3$, transverse smoothing width $\sigma_\perp=1.5\,\mathrm{fm}$, and decoupling temperature $T_{\mathrm{dec}}=160\,\mathrm{MeV}$. At each $b_0$, the BBP/ILE and Liu--Yin prescriptions are evaluated on the same fluid fields and decoupling hypersurface with identical momentum acceptance, $|y|<0.5$ and $0.5<p_T<6\,\mathrm{GeV}/c$. The Vorticity and Shear columns give the corresponding prescription-specific contributions. All polarization entries are in units of $10^{-3}$ and are evaluated with our calculated spectrum and signed Cooper--Frye flux.}\label{tab:prescriptions}
\setlength{\tabcolsep}{5pt}
\begin{tabular}{@{}rlrrr@{}}
\toprule
$b_0$ & Prescription & Vorticity & Shear & Total \\
\midrule
$0.77$ & BBP/ILE & $-2.721$ & $+4.804$ & $+2.083$ \\
$0.77$ & Liu--Yin & $-7.348$ & $+2.843$ & $-4.505$ \\
$0.60$ & BBP/ILE & $-3.764$ & $+5.045$ & $+1.281$ \\
$0.60$ & Liu--Yin & $-8.436$ & $+2.376$ & $-6.060$ \\
\bottomrule
\end{tabular}
\end{table*}

\subsection{Polarization-prescription dependence}

\label{sec:prescription_results}

The preceding comparisons use the ILE expression. To isolate polarization-formula dependence, we evaluate the Liu--Yin expression on the same fluid fields and decoupling hypersurfaces used for the ILE calculation at $\sigma_\perp=1.5\,\mathrm{fm}$, with identical momentum acceptance. Table~\ref{tab:prescriptions} shows that LY reverses the integrated sign at both impact parameters because it gives a more negative thermal-vorticity contribution and a smaller positive shear contribution.

Two formula differences produce the sign reversal. First, the Liu--Yin thermal-vorticity term retains a temperature-gradient contribution because a constant-temperature hypersurface does not require the normal derivative $\partial_\mu T$ to vanish. Second, its shear contribution uses a different tensor projection and energy denominator. For the $b_0=0.77$ reference state, the temperature-gradient part accounts for approximately $70\%$ of the ILE--LY difference, while the different shear terms account for the remaining $30\%$. Both prescriptions include shear-induced polarization. Their opposite signs show that the vorticity--shear balance must be interpreted together with the chosen polarization formula.

\section{Discussion}

\label{sec:discussion}

\subsection{Decoupling-temperature robustness and model sensitivity}

The principal results above use $T_{\mathrm{dec}}=160\,\mathrm{MeV}$. To determine whether the positive sign depends on this decoupling temperature $T_{\mathrm{dec}}$, we repeated the calculation at several temperatures. The positive total ILE harmonic and the shear-dominated ordering persist at $T_{\mathrm{dec}}=157.5$, $158.5$, $160$, and $170\,\mathrm{MeV}$. The competition ratio remains $R\simeq1.53$ from $157.5$ to $160\,\mathrm{MeV}$ and increases to $R=1.66$ at $170\,\mathrm{MeV}$, showing that the positive sign is not tied to a particular decoupling isotherm. Differences between the temperature-scan reconstruction and the primary calculation, together with residual hot cells in the final low-temperature snapshots, affect the detailed temperature dependence but not the sign or component ordering.

The numerical checks of the surface reconstruction and signed Cooper--Frye weighting likewise leave the signs and relative ordering of the kinematic-vorticity contribution $P_{z,s2}^{\omega}$ and the kinematic-shear contribution $P_{z,s2}^{\Xi}$ unchanged.

The absence of transverse expansion before $t_0$ is a separate model limitation. An exploratory radial-flow initialization raises $\langle p_T\rangle$ from $0.9$ to $1.2\,\mathrm{GeV}/c$, reduces the magnitude of the negative kinematic-vorticity contribution $P_{z,s2}^{\omega}$, and strengthens the positive kinematic-shear contribution $P_{z,s2}^{\Xi}$, increasing the total ILE harmonic to about $6.3\times10^{-3}$. This response indicates that longitudinal polarization is sensitive to the transverse velocity field before hydrodynamic evolution. Because the same initialization gives an elliptic flow $v_2=-0.006$, a realistic pre-equilibrium description should be constrained jointly by the momentum spectrum, elliptic flow $v_2$, and longitudinal polarization.

\subsection{Kinematic shear, ideal hydrodynamics, and bulk viscosity}

The kinematic-shear contribution $P_{z,s2}^{\Xi}$ in the ILE formula is determined by velocity gradients on the decoupling hypersurface. Therefore, viscosity can in principle modify $P_{z,s2}^{\Xi}$ indirectly by changing the velocity field, its gradients, and the decoupling hypersurface. The indirect effect of viscosity on the longitudinal-polarization harmonic $P_{z,s2}$ was studied by Palermo et al. using the same ILE polarization formula on a superMC hydrodynamic background \cite{ref52}. In that calculation, introducing bulk viscosity changes the longitudinal harmonic from negative to positive. By contrast, in our calculations with the same ILE polarization formula on an ideal PICR hydrodynamic background, the harmonic is already shear dominated.

Therefore, a positive, shear-dominated harmonic is not by itself evidence for bulk-viscous dynamics: it can also emerge from the velocity gradients and decoupling hypersurface generated by ideal hydrodynamics. Separating the effect of viscosity requires matched initial energy-density and velocity fields, equation of state, and decoupling treatment.

\FloatBarrier
\section{Conclusions}

We have shown that the transverse initial energy-density distribution controls longitudinal $\Lambda$ polarization by changing the competition between the negative kinematic-vorticity contribution $P_{z,s2}^{\omega}$ and the positive kinematic-shear contribution $P_{z,s2}^{\Xi}$. Increasing the initial energy-density normalization $e_{\mathrm{peak}}$ can preferentially reduce $P_{z,s2}^{\Xi}$ and reverse the total ILE harmonic. At fixed $e_{\mathrm{peak}}$, reducing the transverse smoothing width $\sigma_\perp$ generally strengthens both contributions.

Furthermore, the transverse smoothing width $\sigma_\perp$ determines which polarization contribution carries the impact-parameter dependence. At transverse smoothing width $\sigma_\perp=1.0\,\mathrm{fm}$, the impact-parameter dependence is associated mainly with the kinematic-shear contribution $P_{z,s2}^{\Xi}$ and the change in initial eccentricity $\epsilon_{2,\mathrm{init}}$. At $\sigma_\perp=1.5\,\mathrm{fm}$, it is associated mainly with the kinematic-vorticity contribution $P_{z,s2}^{\omega}$ and the change in elliptic flow $v_2$. Moreover, changing the initial energy-density normalization $e_{\mathrm{peak}}$ shows that neither $\epsilon_{2,\mathrm{init}}$ nor $v_2$ alone describes both contributions. The initial-state dependence therefore involves the hydrodynamic evolution of the velocity-gradient fields and the resulting decoupling hypersurface, beyond the information contained in $\epsilon_{2,\mathrm{init}}$ or $v_2$ alone.

Our calculated longitudinal-polarization harmonics obtained with the BBP prescription at isothermal local equilibrium have the same sign and order of magnitude as the ALICE measurements for peripheral Pb--Pb collisions in the $50$--$70\%$ centrality interval. This agreement indicates that the initial-state-controlled, shear-dominated balance at decoupling can account for the measured positive harmonic. By contrast, the Liu--Yin prescription reverses the total sign on the same hydrodynamic backgrounds, showing that the vorticity--shear balance depends on both the hydrodynamic evolution and the polarization formula.

\section*{Acknowledgements}

We thank Pasi Huovinen and Hannu Holopainen for making their Cornelius++ implementation publicly available \cite{ref5}.
\section*{Declarations}
\textbf{Funding:} The work of Y. L. Xie is supported by the National Natural Science Foundation of China under Grant No. 12005196.

\textbf{Conflict of interest:} The authors declare that they have no conflict of interest.

\textbf{Data availability:} The datasets generated and/or analyzed during the current study are available from the corresponding author on reasonable request.

\textbf{Code availability:} This manuscript has no associated code or software deposited in a public repository.
\begingroup

\renewcommand{\bibfont}{\small}

\setlength{\bibsep}{1pt}

\endgroup
\end{document}